\documentclass[sigconf]{acmart}

\AtBeginDocument{%
  }
    
\usepackage{listings}
\usepackage{microtype}
\usepackage{graphicx}
\usepackage{subfigure}
\usepackage{booktabs} %
\usepackage{multirow}
\usepackage{threeparttable}
\usepackage{hyperref}

\usepackage{latexsym}
\usepackage[T1]{fontenc}
\usepackage[utf8]{inputenc}

\usepackage{amsmath}
\usepackage{mathtools}
\usepackage{amsthm}
\usepackage{enumitem}
\usepackage{xspace}
\usepackage{soul}
\usepackage{anyfontsize}
\usepackage[capitalize,noabbrev]{cleveref}

\usepackage{array}
\usepackage{caption}

\usepackage{listings}
\usepackage{textcomp}

\newfloat{Code}{htbp}{Code}

\setcopyright{acmlicensed}
\copyrightyear{2018}
\acmYear{2018}
\acmDOI{XXXXXXX.XXXXXXX}

\acmConference[Conference acronym 'XX]{Make sure to enter the correct
  conference title from your rights confirmation emai}{June 03--05,
  2018}{Woodstock, NY}
\acmISBN{978-1-4503-XXXX-X/18/06}

\begin{document}

\title{SimCE: Simplifying  Cross-Entropy Loss for \\ Collaborative Filtering}

\author{Xiaodong Yang}
\email{xiaodyan@visa.com}
\affiliation{%
  \institution{Visa Research}
  \city{Foster City}
  \country{USA}}

\author{Huiyuan Chen}
\email{hchen@visa.com}
\affiliation{%
  \institution{Visa Research}
  \city{Foster City}
  \country{USA}}

\author{Yuchen Yan}
\email{yucheny5@illinois.edu}
\affiliation{%
  \institution{
University of Illinois Urbana-Champaign}
  \city{Illinois}
  \country{USA}}

\author{Yuxin Tang}
\email{Yuxin.Tang@rice.edu}
\affiliation{%
  \institution{Rice University}
  \city{Houston}
  \country{USA}}

\author{Yuying Zhao}
\email{yuying.zhao@vanderbilt.edu}
\affiliation{%
  \institution{Vanderbilt University}
  \city{Nashville}
  \country{USA}}

\author{Eric Xu}
\email{exu@utexas.edu}
\affiliation{%
  \institution{University of Texas at Austin}
  \city{Austin}
  \country{USA}}

\author{Yiwei Cai}
\email{yicai@visa.com}
\affiliation{%
  \institution{Visa Research}
  \city{Foster City}
  \country{USA}}

\author{Hanghang Tong}
\email{htong@illinois.edu}
\affiliation{%
  \institution{
University of Illinois Urbana-Champaign}
  \city{Illinois}
  \country{USA}}

\renewcommand{\shortauthors}{Xiaodong Yang et al.}


\begin{abstract}
The learning objective is integral to collaborative filtering systems, where the Bayesian Personalized Ranking (BPR) loss is widely used for learning informative backbones. However, BPR often experiences slow convergence and suboptimal local optima, partially because it only considers one negative item for each positive item, neglecting the potential impacts of other unobserved items. To address this issue, the recently proposed Sampled Softmax Cross-Entropy (SSM) compares one positive sample with multiple negative samples, leading to better performance. Our comprehensive experiments confirm that recommender systems consistently benefit from multiple negative samples during  training. Furthermore, we introduce a \underline{Sim}plified Sampled Softmax \underline{C}ross-\underline{E}ntropy Loss (SimCE), which simplifies the SSM using its upper bound. Our validation on 12 benchmark datasets, using both MF and LightGCN backbones, shows that SimCE significantly outperforms both BPR and SSM. 
\end{abstract}



\maketitle

\section{Introduction}
Owing to the superior ability of mitigating information overload, recommender systems have been widely applied in various domains~\cite{koren2009matrix}, including video, news, and e-commerce~\cite{davidson2010youtube,schafer2001commerce}. The primary goal of recommender system is to learn users' preference and then recommend items that match users' interests. To achieve this goal, many techniques have been developed. Among them, one of the most prominent techniques is Collaborative Filtering (CF)~\cite{koren2009matrix,rendle2009bpr}. This technique predicts users' future preference by analyzing the collaborative information between the user and interacted items in their history context, such as purchases and clicks. 

 In generally, CF-based models project each user and item into unique embedding vectors and then compute the user-item preference score in the embedding space~\cite{rendle2010factorization}. Various architectures have been designed to learn these embeddings such that they encode the collaborative information between users and their interacted items and are effective for recommendation. Over the years, the progress of this technique has been notable, transitioning from simple Factorization Machines~\cite{rendle2010factorization} to Deep Neural Networks~\cite{guo2017deepfm}, such as Graph Neural Networks~\cite{he2020lightgcn,chen2021structured,wang2022improving,zhao2024can},  Multi-layer Perceptrons ~\cite{covington2016deep}, Attentive CF~\cite{chen2017attentive}, Autoencoders~\cite{liang2018variational}, and Transformers~\cite{DBLP:conf/cikm/SunLWPLOJ19,chen2022denoising}. Because of their simplicity and efficacy, CF-based approaches have achieved enormous success in practice.

Despite the tremendous efforts in developing more powerful encoders to capture the collaborative signals, the recommendation performance is largely affected by the loss function  in the training~\cite{park2023toward,chen2024towards,wang2022towards,yan2023reconciling,jin2024llm,yeh2023toward}. There are three major categories of loss functions, namely, \textit{pointwise}, \textit{pairwise}, and \textit{listwise} loss. \textit{Pointwise loss} considers each user-item interaction independently and models the recommendation task as either binary classification (e.g., like or not like) or regression problem (e.g., the rating a user gives to an item). The representative losses are therefore Binary Cross-Entropy~\cite{he2017neural} and Mean Square Error loss~\cite{hu2008collaborative, chen2020efficient}. \textit{Pairwise loss} considers the order of preferences for a user over a pair of items. Specifically, it learns to rank interacted items  higher than unobserved items. One of representative methods is Bayesian Personalized Ranking (BPR) loss~\cite{rendle2009bpr} which is derived from the maximum posterior estimator and directly optimizes models for personalized ranking such that the probability of the positive items are higher than those of negative items. Another method is Pairwise Hinge Loss~\cite{hsieh2017collaborative}, also referred as a Max-Margin loss. Given a pair of positive and negative interactions, this function learns to score the positive item higher than negative item by a margin.
\textit{Listwise loss} considers the preferences for a user over a list of items. The Softmax Cross-Entropy loss~\cite{covington2016deep} optimizes probabilities of observed items over other items in a normalized distribution. Due to the heavy computation for all items that limits the usage in large-scale datasets, recently, a simplified version called Sampled Softmax Cross-Entropy (SSM)~\cite{wu2024effectiveness} is further proposed based on sampled negative items instead of all items, alleviating the computation burden. Empirical results show that  SSM achieves state-of-the-art performance when compared to existing loss functions.

In this paper, we further enhance the SSM by proposing a novel loss function named \underline{Sim}plified Sampled Softmax \underline{C}ross-\underline{E}ntropy loss (SimCE), which integrates upper bound optimization. The core strategy of SimCE lies in optimizing the upper bound function derived from the original SSM loss function.
This approach transforms a complex problem into a series of simpler sub-problems, thereby enabling more feasible and effective solutions. By simplifying the optimization, SimCE not only becomes more computationally efficient but also more scalable in handling a variety of complex optimization problems and large datasets, such as large-scale graph neural embedding. To verify the effectiveness of SimCE, we conduct comprehensive experiments on 12 benchmark datasets, using both MF~\cite{koren2009matrix} and LightGCN~\cite{he2020lightgcn} backbones. The experimental results show that SimCE consistently and significantly outperforms both BPR and SSM in terms of the recommendation performance and the training efficiency. Detailedly, among 96 empirical comparisons (i.e., 12 datasets, 2 model types, and 4 metrics), models trained with our SimCE consistently outperform BPR and SSM in 93 instances (i.e., $96.88\%$), where our method significantly surpasses the baselines in most cases with the maximum improvement up to $68.72\%$. In summary, our main contributions are as follows:
\begin{itemize}
    \item We highlight the significance of loss functions and  empirically verify that CF models consistently
benefits from multiple negative samples (SSM v.s. BPR) during the training. 
    \item We propose a Simplified Sampled Softmax Cross-Entropy loss (SimCE) by simplifying the SSM with its upper bound.  Our SimCE can be seamlessly integrated into existing frameworks, offering flexibility and ease of implementation.
    \item We verify the effectiveness and efficiency of SimCE on 12 benchmark datasets over 2 backbones. The experimental results shows that SimCE consistently and significantly outperforms both BPR and SSM in terms of performance.
\end{itemize}

\section{Related Work}
\subsection{Collaborative Filtering}
Collaborative Filtering (CF) aims to predict the user preferences for unobserved items using the collaborative information in the historical user-item interactions~\cite{chen2022graph,koren2009matrix,lai2023enhancing}.  Matrix Factorization (MF)~\cite{koren2009matrix} was firstly proposed to learn to decompose the interaction matrix into two lower-dimensional matrices, representing the latent features of users and items, respectively. The high capability of non-linear analysis by deep neural networks has inspired deep recommender systems to learn and understand the complex patterns in the interactions, including NCF~\cite{he2017neural}, DeepFM~\cite{guo2017deepfm}, DIN~\cite{zhou2018deep}, etc. Additionally, since interaction data can be naturally modeled as graphs where a node represents a user or an item and an edge represents the user-item interaction, recent years have witnessed the development of Graph Neural Network (GNN)-based methods. These methods effectively capture the higher-order collaborative information through message passing~\cite{wang2019neuralngcf,yeh2022embedding,he2020lightgcn,chen2021structured,mao2021ultragcn,yan2023from}. For instance, NGCF~\cite{wang2019neuralngcf} proposes an embedding propagation layer to update node embeddings through neighborhood aggregation. LightGCN~\cite{he2020lightgcn} simplifies the design of NGCF by removing feature transformation and nonlinear activation. UltraGCN~\cite{mao2021ultragcn} further skips infinite layers of message passing to improve efficiency for large-scale datasets.

\subsection{Loss Functions}


Existing loss functions for CF mainly fall into three categories: \textit{pointwise loss}, \textit{pairwise loss}, and \textit{listwise loss}~\cite{park2023toward,chen2024towards,wang2022towards,wang2023federated,nie2022inadequacy}. At a high level, pointwise loss optimizes the user-item relationship for each item independently, pairwise loss considers a pair of positive and negative items simultaneously, and listwise loss takes a list of items into account. More specifically, pointwise optimization treats recommendation as a regression or classification problem and uses losses such as Binary Cross-Entropy~\cite{he2017neural} and Mean Square Error~\cite{hu2008collaborative, chen2020efficient} for optimization. Since the items are optimized independently, pointwise loss often ignores the contexts of other items. Pairwise methods mitigate this issue by incorporating pairs of interactions in the training process. The task becomes learning to score one item over another in terms of the user preference,  ultimately creating a personalized ranking for the user. Representative pairwise methods includes BPR~\cite{rendle2009bpr} and CML~\cite{hsieh2017collaborative}. Listwise optimization, unlike pointwise or pairwise methods, considers all items (or a list of items) directly~\cite{covington2016deep,mao2021simplex,wu2024effectiveness}. The common choice here is the Softmax method~\cite{covington2016deep}, which maximizes the probability of observed items over all others in a normalized distribution. Owing to the ability of considering the impact of other unobserved items, listwise optimization has been proved to be effective in better recommendations compared with pointwise and pairwise methods~\cite{park2023toward}. However, its high computational complexity hinders its wide application in industries, where the number of users and items can be millions or more. Addressing the computational challenges, Cosine Contrastive loss~\cite{mao2021simplex} is proposed to maximize the cosine similarity between user and a positive sample and minimize the similarity between user and multiple negative samples below a given margin. More recently, the Sampled Softmax Cross-Entropy loss (SSM) has been introduced~\cite{wu2024effectiveness}, which approximates the full softmax loss by considering only a sampled subset of negative items, significantly reducing computational complexity~\cite{huang2021mixgcf,klenitskiy2023turning}. 
Our study aligns with this direction, as we aim to further simplify the Sampled Softmax Cross-Entropy loss by utilizing its upper bound.

\section{Methodology}

\begin{figure}
\caption{PyTorch style pseudo-code for three loss functions: BPR, SSM and SimCE.}
\begin{minipage}{0.96\linewidth}
\begin{lstlisting}[language=python,mathescape]
def bpr(user_emb, pos_item_emb, neg_item_emb, gamma=1e-5):
    # user_emb: [batch, dim]
    # pos_item_emb: [batch, dim]
    # neg_item_emb: [batch, dim]
    
    pos_score = torch.mul(user_emb, pos_item_emb).sum(dim=1)
    neg_score = torch.mul(user_emb, neg_item_emb).sum(dim=1)
    loss = -torch.log(gamma + torch.sigmoid(pos_score - neg_score))
    
    return torch.mean(loss)

    
def ssm(user_emb, pos_item_emb, neg_item_emb):
    # user_emb: [batch, dim]
    # pos_item_emb: [batch, dim]
    # neg_item_emb: [batch, num_neg, dim]
    
    num_neg, dim  = neg_item_emb.shape[1], neg_item_emb.shape[2]
    neg_item_emb = neg_item_emb.reshape(-1, num_neg, dim)
    pos_score = torch.mul(user_emb, pos_item_emb).sum(dim=1)
    neg_score = torch.mul(user_emb.unsqueeze(dim=1), neg_item_emb).sum(dim=-1)
    loss = torch.log(1 + torch.exp(neg_score - pos_score.unsqueeze(dim=1)).sum(dim=1))
    
    return torch.mean(loss)

    
def simce(user_emb, pos_item_emb, neg_item_emb, margin=5.0):
    # user_emb: [batch, dim]
    # pos_item_emb: [batch, dim]
    # neg_item_emb: [batch, num_neg, dim]
    
    num_neg, dim  = neg_item_emb.shape[1], neg_item_emb.shape[2]
    neg_item_emb = neg_item_emb.reshape(-1, num_neg, dim)
    pos_score = torch.mul(user_emb, pos_item_emb).sum(dim=1)
    neg_score = torch.mul(user_emb.unsqueeze(dim=1), neg_item_emb).sum(dim=-1)
    neg_score = torch.max(neg_score, dim=-1).values
    loss = torch.relu(margin - pos_score + neg_score)
    
    return torch.mean(loss)
\end{lstlisting}
 \end{minipage}
\end{figure}
\subsection{Task Description}
Let $\mathcal{U}$ and $\mathcal{I}$ denote the sets of users and items, respectively. Given a set of observed user-item interactions $\mathcal{O} = \{(u, i) \mid u \text{ interacted with } i\}$, collaborative filtering  methods aim to predict the score $s(u, i) \in \mathbb{R}$ for each unobserved user-item pair, indicating how likely user $u$ is to interact with item $i$. Based on these predictions, items with the highest scores for each user will be recommended.

In general, most CF methods use an encoder network $f(\cdot)$ that maps each user and item into a low-dimensional representation:
\begin{equation}
    \mathbf{e}_u = f(u), \qquad \mathbf{e}_i = f(i),
\end{equation}
where $\mathbf{e}_u, \mathbf{e}_i \in \mathbb{R}^d$ are the embeddings of user $u$ and item $i$, respectively, with $d$ being the dimension of the embeddings. In practice, the backbone network $f(\cdot)$ can be any model, such as matrix factorization models~\cite{rendle2009bpr} or graph neural networks~\cite{he2020lightgcn}.
Then, the predicted score is defined as the similarity between the user and item representation (e.g., dot product: $s(u, i)=f(u)^Tf(i)$).

\subsection{BPR and SSM}
\subsubsection{\textbf{Bayesian Personalized Ranking (BPR)}}
Regarding the learning objective, most studies adopt the pairwise Bayesian Personalized Ranking (BPR) loss~\cite{rendle2009bpr} to train the model by minimizing:
\begin{equation}
    \mathcal{L}_{BPR} = -\dfrac{1}{|\mathcal{O}|}\sum_{(u,i)\in\mathcal{O}}\log\left[{\sigma}\left(s(u,i) - s(u,j)\right)\right],
    \label{bpr}
\end{equation}
where $\sigma(\cdot)$ is the sigmoid function, and  $j$ is a randomly sampled negative item that the user $u$ has not interacted with. Generally, the BPR function aims to optimize the probability that the target item receives a higher score than a random negative item.

Although yielding promising progress, BPR often suffers from slow convergence and poor local optima~\cite{rendle2009bpr}, partially due to that the BPR employs only one negative example while not considering the potential impacts of  the other negative samples~\cite{sohn2016improved}. 
\subsubsection{\textbf{Sampled Softmax Cross-Entropy Loss (SSM)}} 
To address the problem of BRP, the Softmax Cross-Entropy loss 
computes  a probability distribution over all
items. It then maximizes the probability of the observed items
as compared with that of the unobserved items, that is:
\begin{equation}
    \mathcal{L}_{CE} = -\dfrac{1}{|\mathcal{O}|}\sum_{(u,i)\in\mathcal{O}}\log \frac{\exp(s(u,i))}{\sum_{j=1}^{|\mathcal{I}|}\exp(s(u,j))}.
    \label{ce}
\end{equation}

However, training with the Softmax Cross-Entropy loss over all items becomes computationally prohibitive when scaling to very large-scale datasets that contain million of items. 

To avoid this scalability issue,  the recent proposed Sampled Softmax Cross-Entropy (SSM)~\cite{wu2024effectiveness}  aims to sample a set of negative items for loss
calculation:
\begin{equation}
    \mathcal{L}_{SSM} = -\dfrac{1}{|\mathcal{O}|}\sum_{(u,i)\in\mathcal{O}}\log \frac{\exp(s(u,i))}{\exp(s(u,i)) + \sum_{j=1}^{|\mathcal{N}|}\exp(s(u,j))},
    \label{ssm5}
\end{equation}
where $\mathcal{N}$ is a set of $N$ negative samples sampled for a given positive sample. To gain a deep understanding of SSM, for a positive user-item pair $(u, i)$ in Eq. (\ref{ssm5}), we have the following equivalent:
\begin{equation}
\begin{aligned}
    & -\log \frac{\exp(s(u,i))}{\exp(s(u,i)) + \sum_{j=1}^{|\mathcal{N}|}\exp(s(u,j))} \\
     &= \log(1 +  \sum_{j=1}^{|\mathcal{N}|}\exp(s(u,j) - s(u,i))).
\end{aligned}
    \label{ssm1}
\end{equation}

Clearly, the objective is optimized to identify a positive example from multiple
negative examples. In fact, this idea of simultaneously pushing away multiple negative examples  is not new, and has been widely used in deep metric learning~\cite{sohn2016improved} and recommendation~\cite{huang2021mixgcf}.

\subsection{Our Simplified Sampled Softmax  Cross-Entropy Loss (SimCE)}

Inspired by recent work~\cite{lange2000optimization,mao2023cross}, we propose a Simplified Sampled Softmax  Cross-Entropy Loss (SimCE) by  simplifying the SSM with its upper bound. Give a user $u$, we have:
\begin{equation}
\begin{aligned}
    & -\log \frac{\exp(s(u,i))}{\exp(s(u,i)) + \sum_{j=1}^{|\mathcal{N}|}\exp(s(u,j))} \\
     &= \log\left(1 + \exp(-s(u,i)) \sum_{j=1}^{|\mathcal{N}|}\exp(s(u,j))\right) \\
     & = \log\left(1 + \exp\left(-s(u,i) + \log\sum_{j=1}^{|\mathcal{N}|}\exp(s(u,j))\right)\right) \\
     &\le  \log2 + \max\left( -s(u,i) + \log\sum_{j=1}^{|\mathcal{N}|}\exp(s(u,j)), 0\right) \\
     & \le  \log2 + \max\left( -s(u,i) + \max_{j \in \mathcal{N}}\left(s(u,j)\right) + \log|\mathcal{N}|, 0\right),
\end{aligned}
    \label{SimCE4}
\end{equation}
where the first inequality  holds because $\log(1+\exp(x)) \le \log2 + \max(x,0)$, and the second inequality holds because $\log\sum_i^n\exp(x_i) \le \max(x_1, x_2, \cdots, x_n) + \log n$.

Based on Eq.~\eqref{SimCE4}, our proposed SimCE becomes:
\begin{equation}
    \mathcal{L}_{SimCE} = -\dfrac{1}{|\mathcal{O}|}\sum_{(u,i)\in\mathcal{O}} \max\left( -s(u,i) + \max_{j \in \mathcal{N}}\left(s(u,j)\right) + \gamma, 0\right),
    \label{SimCE}
\end{equation}
here we simply replace $\log|\mathcal{N}|$ with a margin hyperparameter $\gamma$. 
Compared to the SSM in Eq. (\ref{ssm5}), the negative sampling strategy remains unchanged and $|\mathcal{N}|$ negative samples are used in the training. The key difference lies in the optimization/backpropagation process, where our SimCE only selects the \textsl{hardest} negative sample for optimization. Our upper bound minimization indicates that instead of sampling more negative examples, which leads to computational bottlenecks, searching for or generating high-quality negative samples becomes more important for improving model performance and efficiency. \\

\noindent \textbf{Connections among BPR, Hinge loss, and SimCE:}  In the extreme case, if we  set the number of negative samples $|\mathcal{N}| = 1$ during training, our SimCE, which is equivalent to the BPR,  simply degrades to the Hinge loss~\cite{rosasco2004loss,hsieh2017collaborative}:

\begin{equation*}
  \mathcal{L}_{Hinge}=-\dfrac{1}{|\mathcal{O}|}\sum_{(u,i)\in\mathcal{O}} \max( -s(u,i) + s(u,j)+ \gamma, 0).  
\end{equation*}
That is to say, the Hinge loss is actually the upper bound of the BPR loss when the number of negative samples is equal to one. More importantly, our SimCE is a general form of Hinge loss for multiple negative samples.

Finally, we provide a summary of the PyTorch-style pseudo-code for BPR, SSM, and SimCE in Figure 1. Our loss functions can be seamlessly integrated into any existing frameworks, offering flexibility and ease of implementation.

\begin{table}[]
\small
\caption{Statistics of 12 benchmark datasets.}
\label{table1}
\begin{tabular}{lcccc}
    \toprule
Dataset  & \#user & \#item & \#inter. & density \\ \hline
Alibaba  & 106.0k & 53.6k  & 907.5k    & 0.016\% \\   
Pinterest   &55.2k    &9.9k  &1500.8k   &0.274\%  \\
Gowalla   &29.9k   &41.0k   &1027.4k   &0.084\%  \\
iFashion   &300.0k   &81.6k   &1607.8k   &0.007\%  \\
Yelp2018   & 31.7k  & 38.0k  & 1561.4k  & 0.130\%  \\
Douban-Book   & 12.9k  & 22.3k  &598.4k & 0.209\% \\
MovieLens-1M   & 6.0k  & 3.5k  &  575.3k & 2.697\% \\
Amazon-CD &  43.2k &  35.6k  &  777.4k & 0.051\%  \\
Amazon-Book & 52.6k  &  91.6k & 2984.1k  & 0.062\%  \\
Amazon-Beauty   & 22.4k  & 12.1k  & 198.5k   & 0.073\%  \\
Amazon-Kindle & 138.9k  & 98.7k  & 1910.0k  & 0.014\%  \\
Amazon-Movies &  44.4k & 25.9k  & 1070.9k  &  0.096\% \\
\toprule
\label{tab:datasets}
\end{tabular}
\vspace{-15pt}
\end{table}

\setlength{\tabcolsep}{2.1pt}
\setlength{\extrarowheight}{2.1pt}
\begin{table*}[]
\caption{The performance of different loss functions in terms of metrics Recall and NDCG with top 10 and 20 recommendations. \textsl{R} stands for Recall and \textsl{N} stands for NDCG. The \textsl{Improve} indicates the gain of SimCE over SSM.}
\label{tab:results_all}
\begin{tabular}{cc|cccc|cccc|cccc}
\hline
       &         & \multicolumn{4}{c|}{Alibaba} & \multicolumn{4}{c|}{Pinterest} & \multicolumn{4}{c}{Gowalla} \\ \cline{3-14}
       &         & \textbf{R@10}   & \textbf{N@10}   &\textbf{R@20}   & \textbf{N@20}  &\textbf{R@10 }  & \textbf{N@10} & \textbf{R@20}   & \textbf{N@20}  &\textbf{R@10}   & \textbf{N@10} & \textbf{R@20}   & \textbf{N@20}  \\ \hline\hline\multicolumn{1}{c|}{\multirow{4}{*}{\rotatebox[origin=c]{90}{MF}}}       & BPR  & 0.0167& 0.0095& 0.0256& 0.0119& 0.0555& 0.0272& 0.0980& 0.0378& 0.0934& 0.0993& 0.1356& 0.1121   \\ 
\multicolumn{1}{c|}{}     & SSM     & 0.0198& 0.0112& 0.0292& 0.0137& 0.0575& 0.0278& 0.0999& 0.0384& 0.0982& 0.1020& 0.1430& 0.1141   \\ 
\multicolumn{1}{c|}{}     & SimCE    & 0.0324& 0.0189& 0.0447& 0.0215& 0.0683& 0.0341& 0.1158& 0.0460& 0.1179& 0.1261& 0.1693& 0.1413   \\ 
\multicolumn{1}{c|}{}    & Improve & +64.04\%& +68.72\%& +53.13\%& +56.78\%& +18.7\%& +22.72\%& +15.96\%& +19.54\%& +20.12\%& +23.65\%& +18.43\%& +23.87\%   \\ \hline 
\multicolumn{1}{c|}{\multirow{4}{*}{\rotatebox[origin=c]{90}{LightGCN}}} & BPR  & 0.0342& 0.0188& 0.0528& 0.0237& 0.0681& 0.0342& 0.1164& 0.0463& 0.1182& 0.1292& 0.1673& 0.1429   \\ 
\multicolumn{1}{c|}{}     & SSM     & 0.0444& 0.0249& 0.0649& 0.0303& 0.0763& 0.0381& 0.1281& 0.0511& 0.1231& 0.1329& 0.1747& 0.148   \\ 
\multicolumn{1}{c|}{}     & SimCE    & 0.0520& 0.0294& 0.0749& 0.0355& 0.0762& 0.0389& 0.1283& 0.0520& 0.1283& 0.1376& 0.1832& 0.1535   \\ 
\multicolumn{1}{c|}{}     & Improve & +17.23\%& +18.39\%& +15.38\%& +17.09\%& -0.03\%& +2.04\%& +0.13\%& +1.6\%& +4.21\%& +3.53\%& +4.82\%& +3.76\%   \\ \hline 
\hline
       &         & \multicolumn{4}{c|}{iFashion} & \multicolumn{4}{c|}{Yelp2018} & \multicolumn{4}{c}{Douban-Book} \\ \cline{1-14}
\multicolumn{1}{c|}{\multirow{4}{*}{\rotatebox[origin=c]{90}{MF}}}       & BPR  & 0.0346& 0.0193& 0.0528& 0.0241& 0.0267& 0.0302& 0.0468& 0.0377& 0.0831& 0.0918& 0.1271& 0.1019   \\ 
\multicolumn{1}{c|}{}     & SSM     & 0.0367& 0.0204& 0.0559& 0.0256& 0.0290& 0.0327& 0.0504& 0.0408& 0.1065& 0.1195& 0.1525& 0.1280   \\ 
\multicolumn{1}{c|}{}     & SimCE    & 0.0491& 0.0283& 0.0713& 0.0337& 0.0361& 0.0413& 0.0617& 0.0504& 0.1278& 0.1542& 0.1828& 0.1623   \\ 
\multicolumn{1}{c|}{}     & Improve & +33.62\%& +38.55\%& +27.44\%& +31.49\%& +24.24\%& +26.02\%& +22.29\%& +23.63\%& +20.04\%& +28.94\%& +19.87\%& +26.72\%   \\ \hline 
\multicolumn{1}{c|}{\multirow{4}{*}{\rotatebox[origin=c]{90}{LightGCN}}} & BPR  & 0.0578& 0.0328& 0.0864& 0.0404& 0.0343& 0.0397& 0.0588& 0.0487& 0.1015& 0.1169& 0.1506& 0.1261   \\ 
\multicolumn{1}{c|}{}     & SSM     & 0.0722& 0.0418& 0.1048& 0.0507& 0.0399& 0.0456& 0.0673& 0.0554& 0.1299& 0.1574& 0.1828& 0.1641   \\ 
\multicolumn{1}{c|}{}     & SimCE    & 0.0749& 0.0436& 0.1085& 0.0525& 0.0405& 0.0468& 0.0700& 0.0574& 0.1335& 0.1661& 0.1876& 0.1713   \\ 
\multicolumn{1}{c|}{}     & Improve & +3.72\%& +4.23\%& +3.46\%& +3.65\%& +1.45\%& +2.5\%& +4.04\%& +3.66\%& +2.75\%& +5.51\%& +2.6\%& +4.42\%   \\ \hline 
\hline
       &         & \multicolumn{4}{c|}{MovieLens-1M} & \multicolumn{4}{c|}{Amazon-CD} & \multicolumn{4}{c}{Amazon-Book} \\ \cline{1-14}
\multicolumn{1}{c|}{\multirow{4}{*}{\rotatebox[origin=c]{90}{MF}}}       & BPR  & 0.1538& 0.2498& 0.2393& 0.2570& 0.0706& 0.0537& 0.1074& 0.0656& 0.0163& 0.0174& 0.0290& 0.0225   \\ 
\multicolumn{1}{c|}{}     & SSM     & 0.1611& 0.2664& 0.2503& 0.2723& 0.0819& 0.0629& 0.1214& 0.0748& 0.0191& 0.0201& 0.0329& 0.0254   \\ 
\multicolumn{1}{c|}{}     & SimCE    & 0.1850& 0.3173& 0.2794& 0.317& 0.1071& 0.0849& 0.1541& 0.0999& 0.0270& 0.0281& 0.0442& 0.0348   \\ 
\multicolumn{1}{c|}{}     & Improve & +14.85\%& +19.08\%& +11.6\%& +16.41\%& +30.77\%& +35.0\%& +26.95\%& +33.43\%& +41.27\%& +39.49\%& +34.26\%& +36.95\%   \\ \hline 
\multicolumn{1}{c|}{\multirow{4}{*}{\rotatebox[origin=c]{90}{LightGCN}}} & BPR  & 0.1777& 0.3006& 0.2711& 0.3021& 0.0899& 0.0696& 0.1355& 0.0844& 0.0216& 0.0231& 0.0376& 0.0294   \\ 
\multicolumn{1}{c|}{}     & SSM     & 0.1884& 0.3171& 0.2833& 0.3197& 0.1117& 0.0875& 0.1633& 0.1041& 0.0273& 0.0286& 0.0467& 0.0363   \\ 
\multicolumn{1}{c|}{}     & SimCE    & 0.1894& 0.3259& 0.2850& 0.3244& 0.1177& 0.0935& 0.1691& 0.1086& 0.0346& 0.0369& 0.0562& 0.0452   \\ 
\multicolumn{1}{c|}{}     & Improve & +0.52\%& +2.8\%& +0.58\%& +1.45\%& +5.3\%& +6.85\%& +3.56\%& +4.3\%& +26.83\%& +29.2\%& +20.18\%& +24.57\%   \\ \hline 
\hline
       &         & \multicolumn{4}{c|}{Amazon-Beauty} & \multicolumn{4}{c|}{Amazon-Kindle} & \multicolumn{4}{c}{Amazon-Movies} \\ \cline{1-14}
\multicolumn{1}{c|}{\multirow{4}{*}{\rotatebox[origin=c]{90}{MF}}}       & BPR  & 0.0461& 0.0238& 0.0698& 0.0297& 0.1056& 0.0774& 0.1445& 0.0892& 0.0503& 0.0405& 0.0787& 0.0502   \\ 
\multicolumn{1}{c|}{}     & SSM     & 0.0515& 0.0268& 0.0753& 0.0316& 0.1367& 0.1033& 0.1794& 0.1143& 0.0558& 0.0450& 0.0866& 0.0554   \\ 
\multicolumn{1}{c|}{}     & SimCE    & 0.0593& 0.0310& 0.0885& 0.0386& 0.1751& 0.1392& 0.2189& 0.1488& 0.0745& 0.0626& 0.1107& 0.0731   \\ 
\multicolumn{1}{c|}{}     & Improve & +15.06\%& +15.51\%& +17.48\%& +22.27\%& +28.11\%& +34.66\%& +22.04\%& +30.22\%& +33.52\%& +39.1\%& +27.88\%& +31.94\%   \\ \hline 
\multicolumn{1}{c|}{\multirow{4}{*}{\rotatebox[origin=c]{90}{LightGCN}}} & BPR  & 0.0577& 0.0296& 0.0898& 0.0376& 0.1365& 0.1009& 0.1831& 0.1151& 0.0655& 0.0541& 0.1019& 0.0663   \\ 
\multicolumn{1}{c|}{}     & SSM     & 0.0647& 0.0336& 0.0965& 0.0416& 0.1696& 0.1306& 0.2215& 0.1462& 0.0858& 0.0722& 0.1277& 0.0861   \\ 
\multicolumn{1}{c|}{}     & SimCE    & 0.0699& 0.0359& 0.1014& 0.0438& 0.1925& 0.1528& 0.2415& 0.1667& 0.0860& 0.0725& 0.1274& 0.0855   \\ 
\multicolumn{1}{c|}{}     & Improve & +7.95\%& +6.78\%& +5.01\%& +5.34\%& +13.52\%& +16.97\%& +9.05\%& +14.01\%& +0.23\%& +0.42\%& -0.19\%& -0.74\%   \\ \hline 
\end{tabular} 
\end{table*} 

\section{Experiments}

\begin{figure*}
    \centering
    \includegraphics[scale=0.46]{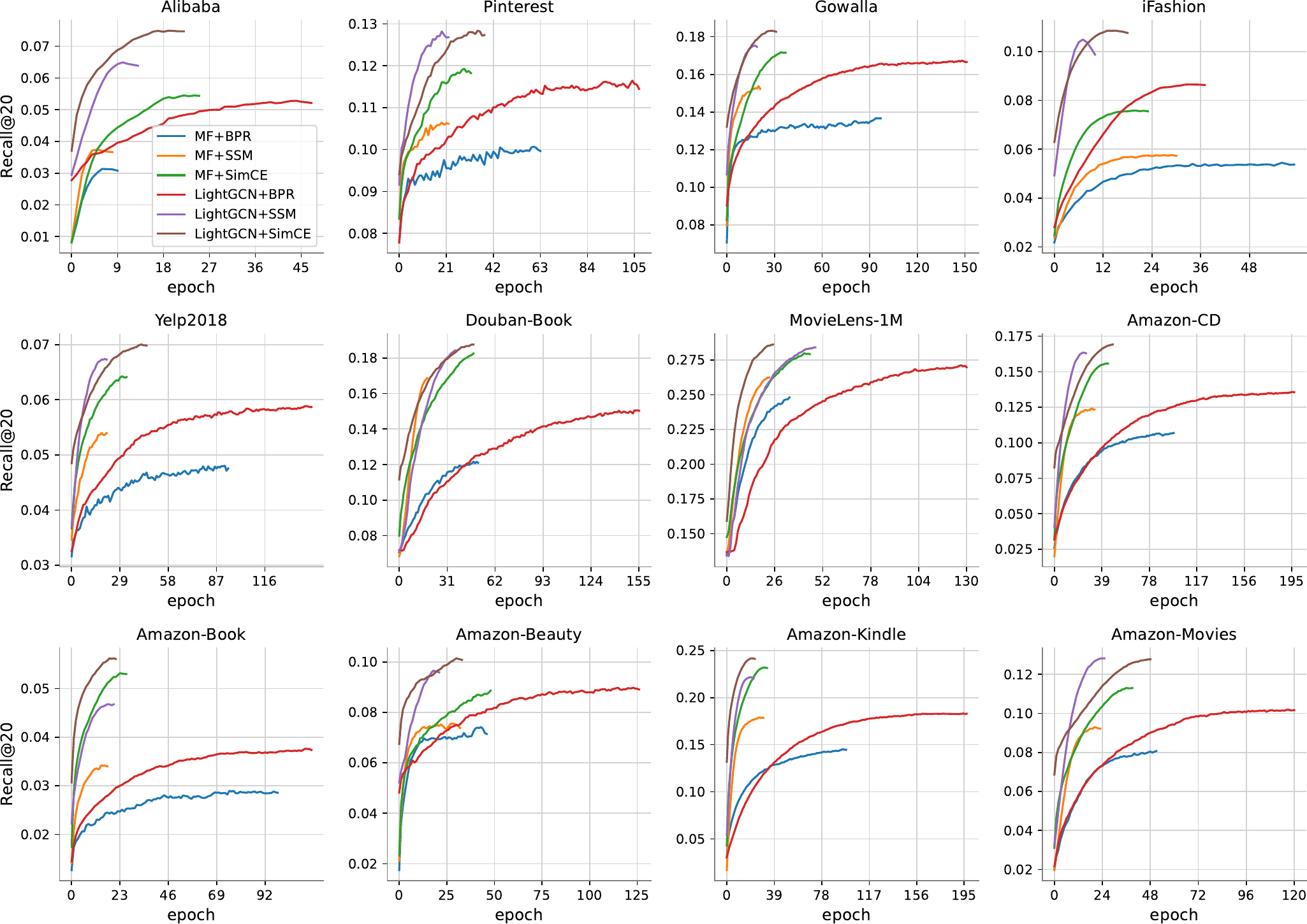}
    \caption{Training curves of different loss functions in terms of Recall@20.}
    \label{fig:Recall@20}
\end{figure*}


\subsection{Setup}

\subsubsection{\textbf{Datasets}} For a thorough evaluation and comparison of different loss functions, we choose 12 commonly used benchmark datasets from various domains, including \textsl{Alibaba}, \textsl{Pinterest}, \textsl{Gowalla}, \textsl{iFashion}, \textsl{Yelp2018}, \textsl{Douban-Book}, \textsl{MovieLens-1M}, \textsl{Amazon-CD}, \textsl{Amazon-Book}, \textsl{Amazon-Beauty}, \textsl{Amazon-Kindle}, and \textsl{Amazon-Movies}. The short description of each dataset are as follows:
\begin{itemize}
    \item \textbf{Alibaba}~\cite{yang2020understanding} is created using data from a large E-commerce platform named Alibaba, which consists of users' purchase history and products with their attribute information. This dataset is commonly constructed as a user-item graph for recommendation.
    \item \textbf{Pinterest}~\cite{he2017neural} is created from a popular social media platform named Pinterest. It contains users' pin records with implicit feedback and is commonly used for evaluating content-based image recommendation. This revised version retains users with at least 20 interactions.
    \item \textbf{Gowalla}\footnote{https://snap.stanford.edu/data/loc-Gowalla.html} is extracted from a social networking website named Gowalla, which contains user's check-in location records. This revised dataset keeps users and items with more than 10 interactions. 
    \item \textbf{iFashion}~\cite{chen2019pog} is constructed based on the data from an application named iFashion, which is used for outfitting recommendation in Taobao, a large online consumer-to-consumer platform. It contains users and outfits interaction records.
    \item \textbf{Yelp2018}\footnote{https://www.yelp.com/dataset} is collected from a business review platform named Yelp. This dataset is an edition used for the 2018 Yelp challenge and contains users and items with more than 10 interactions.
    \item \textbf{Douban-Book}\footnote{https://www.kaggle.com/datasets/fengzhujoey/douban-datasetratingreviewside-information} is collected from one of the most influential cultural community platforms in China, Douban.com. It contains different kinds of raw information, i.e., ratings, reviews, item details, user profiles, tags, and date.
    \item \textbf{MovieLens-1M}\footnote{https://grouplens.org/datasets/movielens/} is a dataset that contains 1 million ratings of users on movies, which is commonly used to evaluate recommendation algorithms.

    \item \textbf{Amazon}\footnote{https://jmcauley.ucsd.edu/data/amazon/} datasets contains records related to user and product reviews across various categories on Amazon.com, In this work, we mainly focus on the categories of CDs, Books, Beauty, Kindle, and Movies.
\end{itemize}

 The detailed statistics of these datasets are provided in Table~\ref{tab:datasets}.  Moreover, we utilize two widely adopted metrics, Recall and Normalized Discounted Cumulative Gain (NDCG)~\cite{he2020lightgcn,wu2024effectiveness}, to assess the performance on top-$k$ recommendation. For our experiments, we set $k$ to 10 and 20 by default. 

\subsubsection{\textbf{Backbones}} To evaluate the flexibility of our loss function, we equip SimCE with two popular CF backbones: Matrix Factorization (MF)~\cite{koren2009matrix} and LightGCN~\cite{he2020lightgcn}:
\begin{itemize}
    \item \textbf{Matrix Factorization (MF)}~\cite{koren2009matrix}: A classic CF model that seeks to use the inner product of user and item embeddings as its preference predictor, which captures the interaction between the user and item features.
    \item \textbf{LightGCN}~\cite{he2020lightgcn}:  A simplified and efficient variant of graph convolutional networks tailored specifically for collaborative filtering, which achieves the promising performance in graph-based CF methods. 
\end{itemize}

\subsubsection{\textbf{Hyperparameter Settings}}
The model architectures of MF and LightGCN and their hyperparameter settings are consistent with those in the original frameworks. Specifically, the embedding dimension of users/items is set to 64, and the number of layers of LightGCN is set to 2 to avoid the over-smoothing issue. A fixed learning rate of $1e^{-4}$ and batch size of $1,024$ are applied across all datasets. We follow the same data splitting strategy to create training, validation, and testing sets in each dataset~\cite{he2020lightgcn,wu2024effectiveness}.

For the loss functions (BPR, SSM, and SimCE), given that BPR is a pairwise loss considering a single negative sample, the number of negative samples for BPR is always set to 1. For both SSM and our SimCE, we search 8 distinct number of negative samples within $|\mathcal{N}| = [4,8,16,32,64,128,256,512]$ during training. This range of negative sampling allows us to thoroughly analyze the impact of negative samples on recommendation performance. Furthermore, our SimCE also has an additional margin hyperparameter $\gamma$ in Eq. (\ref{SimCE}), which we search within $[1.0,5.0,10.0]$ to evaluate the effect of the margin on the performance. Our code and datasets are available: \href{https://github.com/KevinC2015/SimCE}{https://github.com/KevinC2015/SimCE}.

\subsection{Overall Performance}

The  results for all baselines BPR, SSM, and SimCE are shown in Table~\ref{tab:results_all}. With 12 datasets, 2 model types, and 4 metrics, there are a total of 96 comparison instances. We can observe that models trained with our SimCE consistently outperform both BPR and SSM in 93 instances. In most cases, our method significantly surpasses the baselines, with the maximum improvement reaching up to $68.72\%$. The 3 instances where our method does not exceed the baselines including the Pinterest and Amazon-Movies datasets using the LightGCN model. Here, our SimCE only marginally underperforms SSM, making the performance difference negligible. Thus, our method is comparable in these instances as well. 

\begin{figure*}
    \centering
    \includegraphics[scale=0.46]{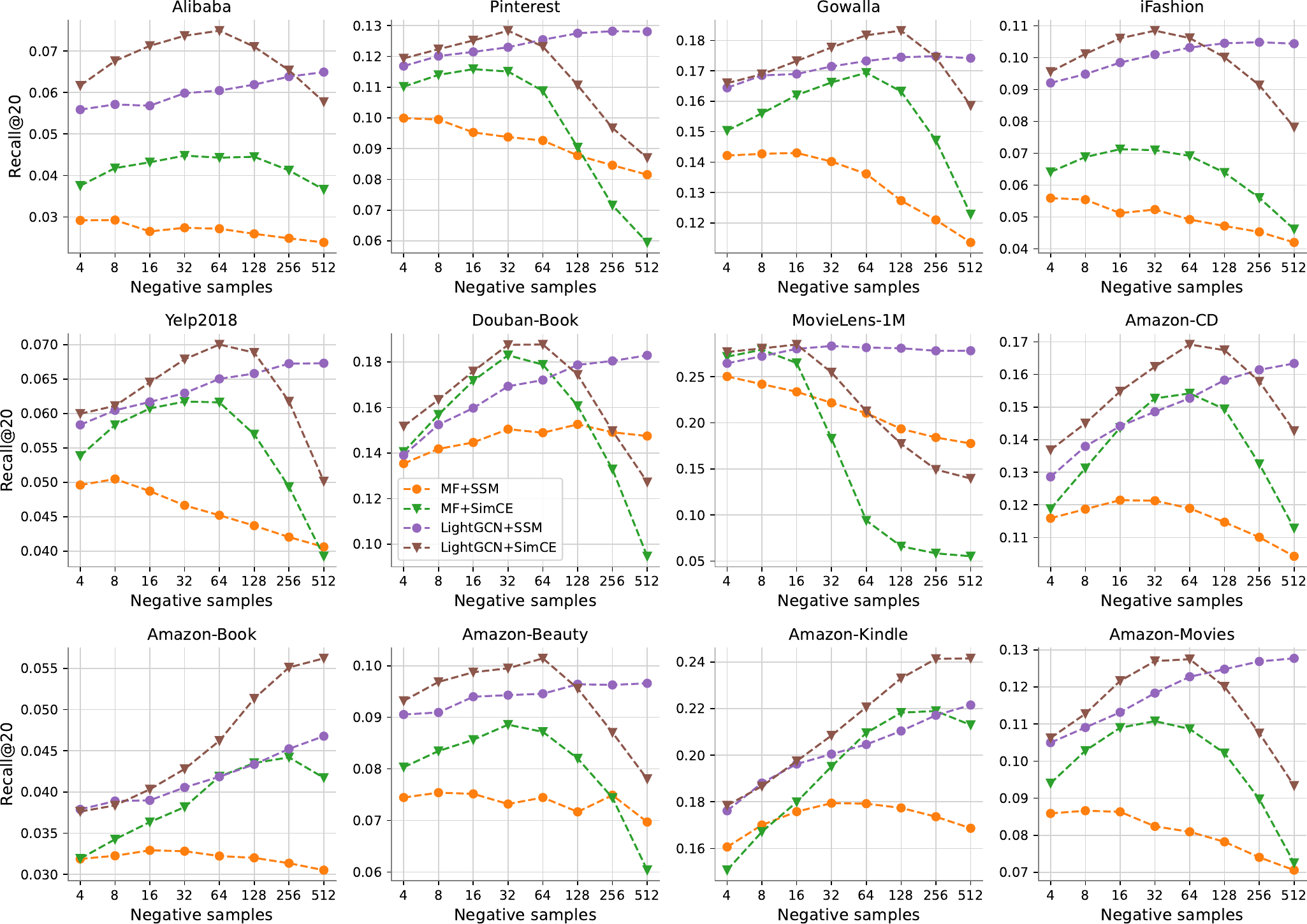}
    \caption{The impact of the number of negative samples $|\mathcal{N}|$ for both SSM and SimCE in terms of  Recall@20.}
    \label{fig:negative-Recall@20}
\end{figure*}

The results also indicate that, compared to listwise losses like SSM and SimCE, the pairwise loss BPR may be trivial for ranking recommendations in practice, given its approach to optimize the ranking of a pair of items, while the listwise loss optimizes over a list of items. It is noteworthy that even though our SimCE only considers the hardest negative {samples} to compute loss values, it also takes the all negative samples into account. That is, it also needs to compute the similarity score for all the negative samples to pinpoint the hardest negative sample for each positive user-item pair. Therefore, our SimCE consistently
benefit{s} from multiple negative samples during the training stage.

Additionally, the performance improvement our SimCE achieves on MF is typically greater than that on LightGCN for most of datasets. This may result from LightGCN's superior ability to capture high-order user-item interactions in recommendations. That implies that both model architectures and loss functions are important in designing the recommender systems.  Moreover, aside from smaller but with higher density datasets like MovieLens-1M, our method also yields considerable improvement on large-scale datasets such as Alibaba, iFashion, and Amazon-Kindle. This underlines SimCE's broad applicability to diverse datasets with different levels of {densities}.

\begin{table*}
 \caption{Training efficiency. Time stands for training time (\textit{sec}) per epoch. C. Epoch stands for Converged Epoch where the best performance occurs.}
\renewcommand{\arraystretch}{0.96}
 \begin{tabular}{cl|cc|cc|cc|cc|cc|cc}
 \hline
    &    & \multicolumn{2}{c|}{Gowalla}& \multicolumn{2}{c|}{iFashion}& \multicolumn{2}{c|}{Yelp2018}& \multicolumn{2}{c|}{Kindle}& \multicolumn{2}{c|}{Book}& \multicolumn{2}{c}{Movies}\\ \cline{3-14} 
    &     & Time & C. Epoch  & Time & C. Epoch  & Time & C. Epoch  & Time & C. Epoch  & Time & C. Epoch  & Time & C. Epoch \\ \hline
 \multicolumn{1}{c|}{\multirow{3}{*}{\rotatebox[origin=c]{90}{MF}}}       & BPR   & 9.0 & 94 & 69.5 & 56 & 10.9 & 91 & 36.7 & 98 & 26.6 & 99 & 10.5 & 48\\ 
 \multicolumn{1}{c|}{}    & SSM  & 272.6 & 9 & 504.2 & 8 & 415.6 & 9 & 534.9 & 11 & 766.1 & 8 & 268.5 & 12\\ 
 \multicolumn{1}{c|}{}     & SimCE  & 269.6 & 21 & 505.6 & 18 & 408.3 & 26 & 593.2 & 28 & 763.2 & 22 & 282.5 & 22\\ 
 \hline
 \multicolumn{1}{c|}{\multirow{3}{*}{\rotatebox[origin=c]{90}{LightGCN}}} & BPR   & 15.9 & 148 & 89.9 & 34 & 27.2 & 141 & 63.4 & 194 & 86.1 & 111 & 19.0 & 117\\ 
 \multicolumn{1}{c|}{}    & SSM  & 294.4 & 15 & 560.7 & 6 & 462.8 & 18 & 570.8 & 19 & 843.7 & 17 & 272.4 & 22\\ 
 \multicolumn{1}{c|}{}    & SimCE  & 290.0 & 16 & 539.2 & 17 & 450.5 & 24 & 582.5 & 20 & 863.2 & 18 & 286.5 & 32\\ 
 \hline
 \end{tabular}
  \label{tab:efficiency}
 \end{table*}


 \begin{figure*}[t!]
 \centering
 \includegraphics[scale=0.52]{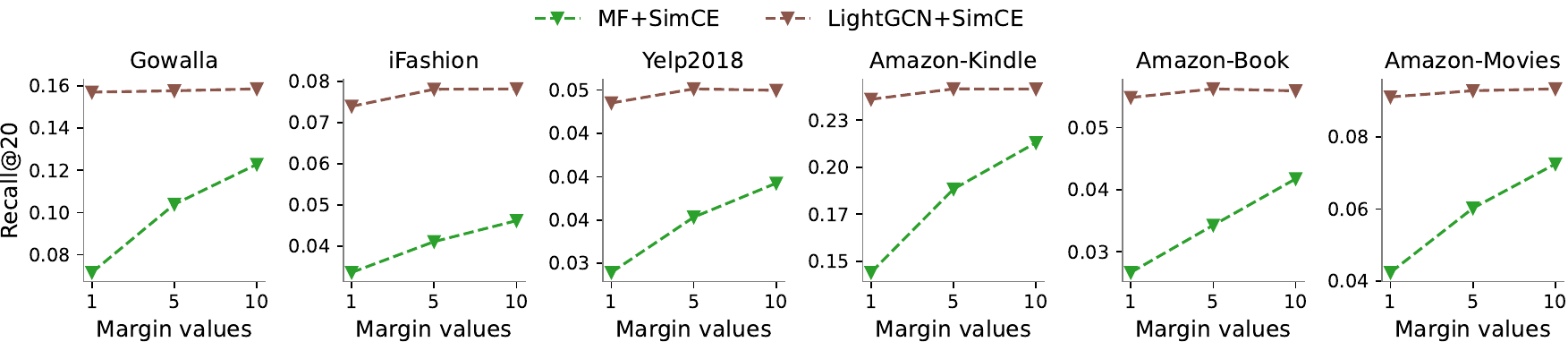}
 \caption{The impact of different margin values $\gamma$ on Gowalla, iFashion, Yelp,  Kindle, Book and Movies datasets.}
 \label{fig:margins}
\end{figure*}

\subsection{Training efficiency}

To assess the training process of various loss functions, we illustrate the training curves for all datasets in terms of Recall@20, as depicted in Figure~\ref{fig:Recall@20}. Regarding the convergence speed, training with our SimCE usually requires $20\sim40$ epochs to attain the peak performance, which is comparable to SSM and  consistently achieve{s} a better performance. On the other hand, training with BPR requires significantly more epochs and but achieves a lower performance.

To obtain a deeper insight into the training efficiency of different loss functions, we compute and compare their training time per epoch and the converged epochs where the best performance is achieved.  We select six large-scale datasets: Gowalla, iFashion, Yelp2018, Amazon-Kindle, Amazon-Book, and Amazon-Movies. We set the number of negative samples to $512$ for both SSM and our SimCE. A larger number of negative samples increases computation, making the efficiency comparison fairer. The overall results of different datasets are shown in Table~\ref{tab:efficiency}. 

In terms of training time per epoch, our SimCE is comparable to SSM for all three large datasets. The training time of BPR is substantially less for one epoch, because it considers only one negative sample, while both our SimCE and SSM take 512 negative samples per positive sample into account during training. 
Regarding the converged epochs, BPR requires more epochs which is consistent with the fact that BPR often experiences slow convergence and suboptimal local optima, partially because it only considers one negative item for each
positive item, neglecting the potential impacts of other unobserved
items. In addition, the results show the converged epochs of our method is slightly greater than SSM. This is because, during the backpropagation of SimCE, only the gradients for the hardest negative sample are updated, whereas SSM updates the gradients for all negative samples but performs worse than our SimCE in term of the accuracy. 

In summary, the experimental results demonstrate the superiority of our proposed SimCE. It generally outperforms all baselines across {12} datasets and has a comparable complexity to other loss functions, especially SSM.

\subsection{The Impact of  Negative Samples}
We analyze the impact of negative samples on SSM and our SimCE performance, using 8 different negative sampling settings. The results for all 12 datasets are shown in Figure~\ref{fig:negative-Recall@20}. In general, we observe that the model performance of both MF and LightGCN initially improves with more negative samples but declines after a certain point. The results indicate that while multiple negative samples are crucial in training, merely increasing the number of negative samples does not always yield better performance.   

An additional observation is that, in terms of the model types MF and LightGCN, the results show the degradation of MF happens at {a} smaller number of negative samples and is more drastic than LightGCN, indicating MF is less robust to  hard negative samples.  
We leave this as one future work to investigate the high-quality negative samples in recommendation.
In practice, for our SimCE, we can choose $\mathcal{N} = 64$ to achieve a good balance between accuracy and computation for large-scale datasets.

\subsection{The Impact of the Margin $\gamma$}
One important hyperparameter of our SimCE is the margin  $\gamma$ in Eq. (\ref{SimCE}), which controls the relative importance of the loss associated with positive and negative samples. We examine its effect using three different values $[1.0, 5.0, 10.0]$ on the six large-scale datasets as selected in Sec 4.4. We observe the similar trend in other datasets and omit their results due to the page limit.

The number of negative samples is set at $512$. The performance is evaluated in terms of Recall@20 and results are displayed in Figure~\ref{fig:margins}.  Intuitively, SimCE is optimized to enhance the similarity between positive pairs while reducing the similarity of negative pairs below a specified margin. As shown in Figure~\ref{fig:margins},  it is apparent that larger $\gamma$ values might enhance performance, with the improvement being particularly noticeable on MF with SimCE. The margin $\gamma$ serves as a similarity function  to the one in the Hinge loss, serving as a threshold for the ReLU function in Eq~\ref{SimCE}, and aiming to minimize the loss by a margin. In contrast, LightGCN with SimCE remains relatively stable across varying margin values. We do not observe further improvement with larger margin values ($\gamma$). This stability suggests that LightGCN with SimCE is less sensitive to the choice of the margin parameter, ensuring consistent performance regardless of the specific value of $\gamma$ chosen.

\section{Conclusion and Future Work}
In this work, we focus on the design of loss functions for recommender systems. While many studies have aimed to improve model architectures, particularly interaction encoders, we emphasize the significant impact of loss functions on recommendation performance. First, we validate that the existing SSM with multiple negative samples generally outperforms the BPR loss, which uses only one negative sample. Building on this, we propose a Generalized Sampled Softmax Cross-Entropy loss (SimCE), which simplifies the SSM loss using its upper bound. Additionally, we provide PyTorch-style pseudo-code for two commonly used loss functions, BPR and SSM, as well as our novel SimCE loss. Extensive experiments on 12 benchmark datasets using MF and LightGCN backbones validate the effectiveness and efficiency of the proposed SimCE loss. 

For future work, we plan to extend our SimCE to other recommendation tasks, such as knowledge graph recommendations and sequential recommendations. We are also interested in exploring the use of generative algorithms to create high-quality negative samples for better training.

\bibliographystyle{ACM-Reference-Format}
\bibliography{sample-base}

\end{document}